\documentclass[sigconf]{acmart}

\AtBeginDocument{%
  \providecommand\BibTeX{{%
    \normalfont B\kern-0.5em{\scshape i\kern-0.25em b}\kern-0.8em\TeX}}}

\copyrightyear{2023}
\acmYear{2023}
\setcopyright{rightsretained}
\acmConference[CHI '23]{Proceedings of the 2023 CHI Conference on Human Factors in Computing Systems}{April 23--28, 2023}{Hamburg, Germany}
\acmBooktitle{Proceedings of the 2023 CHI Conference on Human Factors in Computing Systems (CHI '23), April 23--28, 2023, Hamburg, Germany}\acmDOI{10.1145/3544548.3581392}
\acmISBN{978-1-4503-9421-5/23/04}

\usepackage{xcolor}
\usepackage{multicol}
\usepackage{dblfloatfix}
\usepackage{soul}

\usepackage{color, colortbl}

\definecolor{Gray}{gray}{0.95}

\usepackage{hyperref}
\hypersetup{
pdftitle={Template},
pdfsubject={Human-Computer Interaction},
pdfauthor={},
pdfkeywords={}
}

\definecolor{brown}{rgb}{0.59, 0.29, 0.0}
\definecolor{darkgray}{rgb}{0.59, 0.59, 0.59}
\definecolor{tablegray}{gray}{.9}

\newcommand\revision[1]{\textcolor{blue}}

\usepackage[utf8]{inputenc}
\usepackage{diagbox}
\usepackage{colortbl}

\usepackage{tabularx}

\usepackage{nameref}

\usepackage{xspace}
\usepackage{enumitem}
\usepackage{mathtools}
\usepackage{commath}

\usepackage{amssymb}
\usepackage{pifont}
\newcommand{\cmark}{\ding{51}}%
\newcommand{\xmark}{\ding{53}}%

\newcommand{\customtilde}{{\raise.17ex\hbox{$\scriptstyle\sim$}}}

\newcommand{\etal}{et~al.\xspace}

\begin{document}

\title{IMUPoser: Full-Body Pose Estimation using IMUs in Phones, Watches, and Earbuds}

\author{Vimal Mollyn}
\email{vmollyn@andrew.cmu.edu}
\orcid{0000-0002-9085-8830}
\affiliation{%
  \institution{Carnegie Mellon University}
  \city{Pittsburgh}
  \state{PA}
  \country{USA}
}

\author{Riku Arakawa}
\email{rarakawa@cs.cmu.edu}
\orcid{0000-0001-7868-4754}
\affiliation{
  \institution{Carnegie Mellon University}
  \city{Pittsburgh}
  \state{PA}
  \country{USA}
}

\author{Mayank Goel}
\email{mayank@cs.cmu.edu}
\orcid{0000-0003-1237-7545}
\affiliation{%
  \institution{Carnegie Mellon University}
  \city{Pittsburgh}
  \state{PA}
  \country{USA}
} 

\author{Chris Harrison}
\email{chris.harrison@cs.cmu.edu}
\orcid{0000-0001-5312-3619}
\affiliation{%
  \institution{Carnegie Mellon University}
  \city{Pittsburgh}
  \state{PA}
  \country{USA}
}

\author{Karan Ahuja}
\email{kahuja@cs.cmu.edu}
\orcid{0000-0003-2497-0530}
\affiliation{%
  \institution{Carnegie Mellon University}
  \city{Pittsburgh}
  \state{PA}
  \country{USA}
}

\renewcommand{\shortauthors}{Mollyn et~al.}


\settopmatter{authorsperrow=3}

\begin{abstract}
Tracking body pose on-the-go could have powerful uses in fitness, mobile gaming, context-aware virtual assistants, and rehabilitation. However, users are unlikely to buy and wear special suits or sensor arrays to achieve this end. Instead, in this work, we explore the feasibility of estimating body pose using IMUs already in devices that many users own --- namely smartphones, smartwatches, and earbuds. This approach has several challenges, including noisy data from low-cost commodity IMUs, and the fact that the number of instrumentation points on a user's body is both sparse and in flux. Our pipeline receives whatever subset of IMU data is available, potentially from just a single device, and produces a best-guess pose. To evaluate our model, we created the IMUPoser Dataset, collected from 10 participants wearing or holding off-the-shelf consumer devices and across a variety of activity contexts. We provide a comprehensive evaluation of our system, benchmarking it on both our own and existing IMU datasets. 
\begin{figure*}[b]
    \centering
    \includegraphics[width=\textwidth]{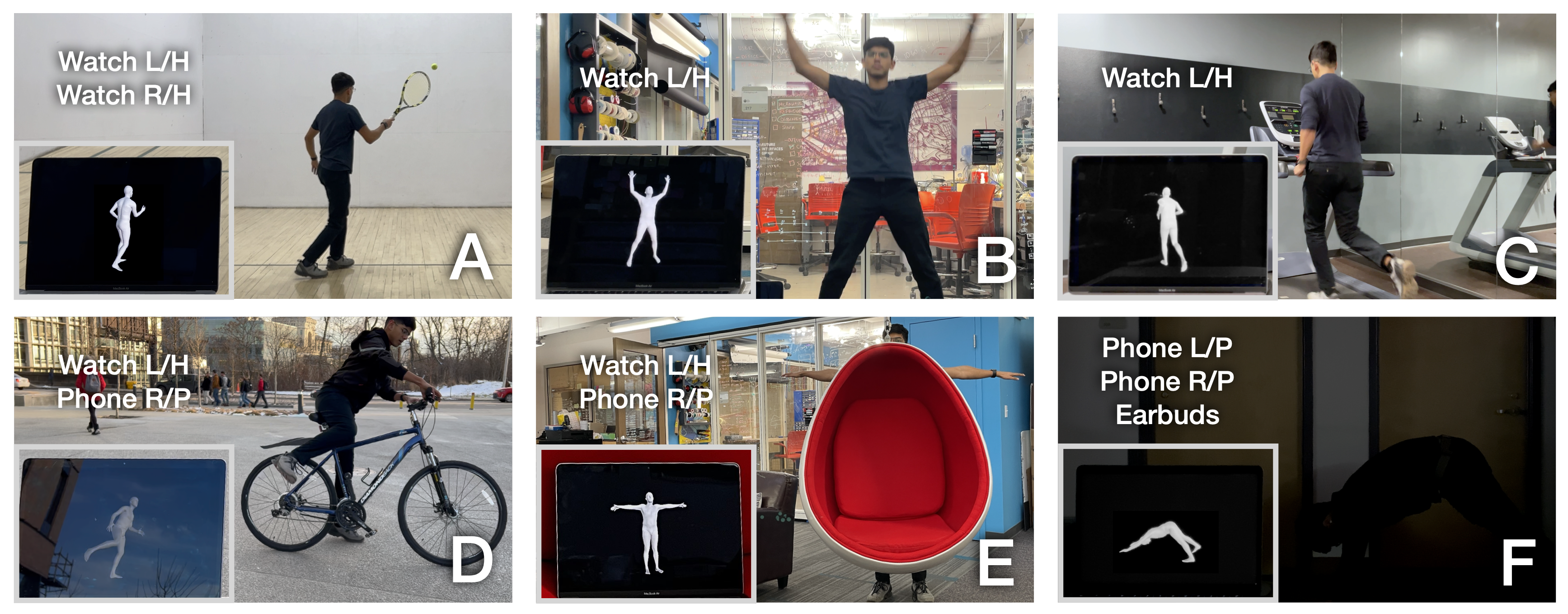}
    \caption{Real-time pose estimation (inset photos) powered by consumer mobile devices (listed in each photo) could have uses across many domains, including sports (A), rehabilitation (B), fitness (C), and transportation (D). Note also that IMUPoser is robust to occlusion (E) and lighting conditions (F). Abbreviation key: L-Left, R-Right, H-Hand, and P-Pocket.}
    \label{fig:example_applications}
\end{figure*}

\end{abstract}





\begin{teaserfigure}
\centering
  \includegraphics[width=0.89\linewidth]{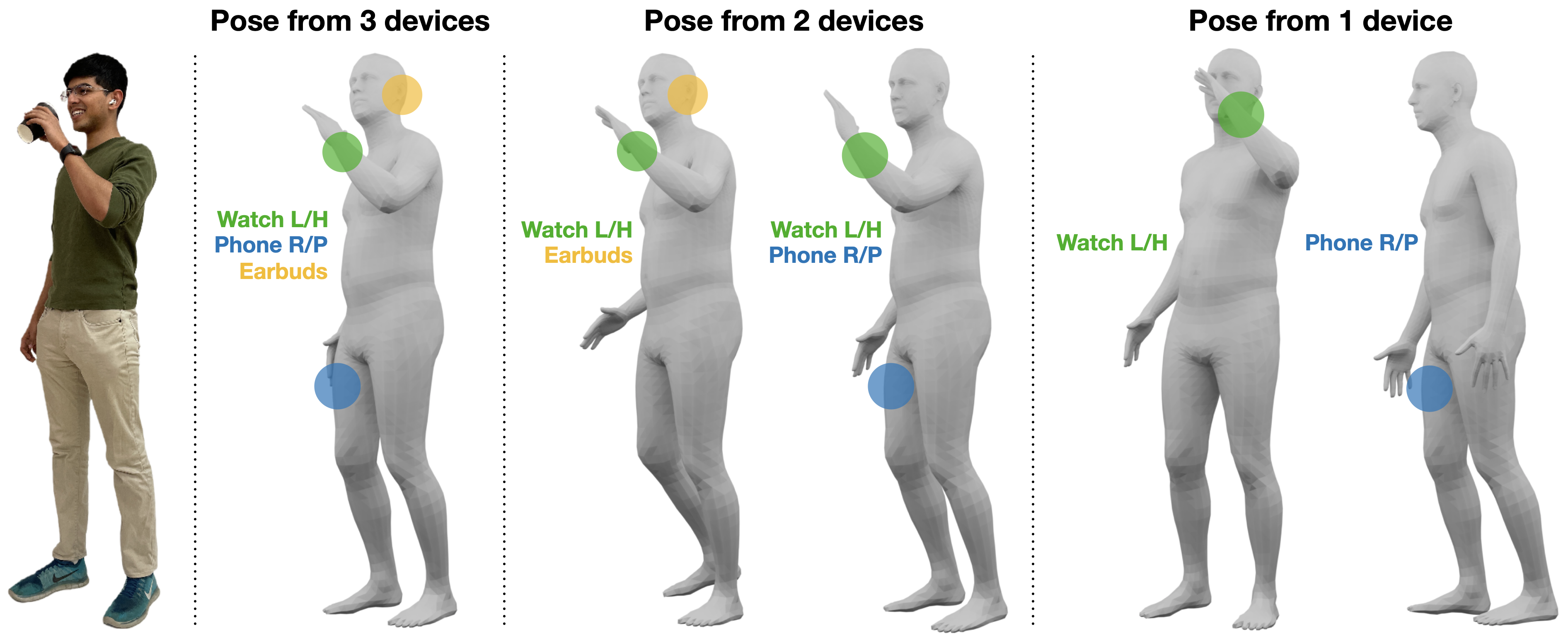}
  \vspace*{-3mm}\caption{Using whatever mobile devices a user has with them, IMUPoser estimates full-body pose. In the best case, a user can have a smartphone, smartwatch and earbuds (pose from 3 devices). Of course, the number of devices will vary over time, e.g., earbud use is intermittent and not everyone wears a smartwatch. This means IMUPoser must track what devices are present, where they are located, and use whatever IMU data is available. Abbreviation key: L-Left, R-Right, H-Hand, and P-Pocket. }\vspace*{1mm}
  \label{fig:teaser}
\end{teaserfigure}

\maketitle

\section{Introduction} 
Full-body motion capture is commonplace in movie visual effects and is slowly entering the consumer realm in areas such as virtual reality. Full-body pose tracking has obvious applications in gaming \cite{Kinect_games}, fitness \cite{khurana2018gymcam}, rehabilitation \cite{mousavi2014review}, life logging \cite{jalal2012depth}, and context-aware interfaces \cite{dov, adeli2020socially}. For example, digital assistants with knowledge of pose could help a football player improve their form, or a patient recovering from surgery monitor changes in their gait. However, at present, most consumers have no tools to track their pose, nor do they want to retrofit sensors into their homes or wear special-purpose suits or accessory devices. However, if it was possible to generate useful pose information from devices users already own, it could have a significant impact. 


Most computing devices we carry with us contain IMUs, most notably smartphones, smartwatches, and wireless earbuds. In this work, we study how we can use this ecosystem of worn and mobile devices to estimate a user's body pose in real-time and with no external infrastructure. This approach introduces new challenges prior sparse IMU pose models (e.g., \cite{von2017sparse, huang2018deep}) have not faced. Uniquely, the position and number of tracked body locations can change on the go. For instance, a user can take a phone from their left pocket into their right hand, or a user can add to the number of sensed points by wearing their earbuds. Our model must accept various combinations of incomplete inputs and gracefully degrade as the number of active devices reduces (potentially to one). Secondarily, our system must work with IMU data received from consumer devices that are noisier than professional-grade motion capture suits (e.g., XSens \cite{xsens}) used in highly-related prior work such as Sparse Internal Poser \cite{von2017sparse}, Deep Inertial Poser \cite{huang2018deep}, and TransPose \cite{yi2021transpose}. Table~\ref{tab:imu-systems-overview} provides an overview of the most related prior work.

To evaluate our method, we created a novel dataset: professional-grade Vicon optical tracking paired with commodity device IMU data from common worn/held locations. Unsurprisingly –- given that we have \textit{at most} three body positions to estimate hundreds of degrees of freedom in the human body -- our pose output is an approximation. However, it is rarely wildly incorrect, and most often, the main gestalt of a user's pose and locomotion mode is captured. This "low-fi" pose output is ill-suited for high-fidelity applications, such as special effects motion capture or virtual reality avatars, where users expect a mostly-faithful body representation. Nonetheless, the adaptive and mobile nature of IMUPoser enables passive and longitudinal sensing of the user (potentially even from a single device), making it especially well-suited for health and wellness applications. For example, this low-fi body tracking could be valuable for boosting accuracy in calorie counting, tracking progress in a physical therapy regime, and monitoring exercise form and rep count. We highlight some example uses in Figure \ref{fig:example_applications}.

\section{Related Work} 

We now review the related work in the area of full-body digitization. We look at both external and worn capture systems, and then review IMU-based pose-sensing approaches that are most related to our work. For an in-depth review of past and current approaches for pose estimation, we refer readers to surveys by Desmarais~\etal\cite{Desmarais2021} and Nguyen~\etal\cite{Nguyen2022}.

\subsection{Body Capture using External Sensors}
\begin{table*}[t]
\centering
\begin{tabular}{l c c c c c }
\toprule
\textbf{System}    & \textbf{\# Inst. Joints}  & \textbf{Sensor FPS (Hz)} & \textbf{Consumer Device} & \textbf{Real-time} & \textbf{MPJVE (cm)}   \\
\midrule

XSens \cite{xsens}  & 17  & 120 & \xmark & \xmark & -       \\
SIP \cite{von2017sparse}  & 6  & 60 & \xmark &  \xmark & 7.71         \\
DIP \cite{huang2018deep}       & 6 & 60 & \xmark & \cmark & 8.96               \\
Transpose \cite{yi2021transpose}  & 6  & 90 & \xmark & \cmark  & 7.09       \\
PIP \cite{yi2022physical}       & 6  & 60 & \xmark & \cmark    & 5.95             \\
Tautges~\etal\cite{tautges2011motion}  & 4  & 25 & \xmark & \xmark & -       \\
\rowcolor{Gray}
IMUPoser (our work)     & 1--3  & 25 & \cmark & \cmark & 12.08             \\
\bottomrule
\end{tabular}
\vspace*{2mm}\caption{Comparison of worn-IMU, full-body pose estimation systems. MPJVE is calculated on the DIP-IMU dataset \cite{huang2018deep}.}
\label{tab:imu-systems-overview}
\end{table*}

There exists a wide range of approaches and solutions to estimate users' pose using external sensors. Commercial systems such as Vicon \cite{Vicon} and OptiTrack \cite{OptiTrack} use specialized hardware, including high-speed infrared cameras that track retroreflective markers attached to users' whole body or individual parts, such as the face or hands. After a calibration procedure, these systems can track large spaces at millimeter accuracy. These approaches are often used for movies, games, and character animation. The cost and setup requirements, however, preclude most consumer applications.

Current approaches for whole-body pose estimation in \textit{consumer} applications typically rely on cheaper sensors and require less calibration.
Depth cameras such as the Microsoft Kinect \cite{Shotton2013, OpenNI, Wei2012} and Intel RealSense \cite{Intel_realSense} provide sufficient pose accuracy using medium-cost sensors for applications in VR and gaming.
Zimmermann~\etal\cite{zimmermann20183d} and Michael~\etal\cite{michel2017markerless}, for example, extend these approaches and combine depth imagery with RGB data for improved accuracy.
Such commercially-available sensors provide a good balance between cost, availability and accuracy, however, they are generally immobile setups. Recent successes in computer vision and deep learning have made it possible to extract pose data from monocular RGB cameras.
This includes approaches that infer 2D pose of one or multiple humans from a single image \cite{Cao:2017, Cao2019,Papandreou:2018},  multiple cameras \cite{DeAguiar2008}, or estimating 3D poses from 2D images \cite{mehta2017vnect}.

There also exists specialized external hardware for pose tracking in VR and AR \cite{Marchand2016}. For example, the HTC Vive \cite{Vive, Vive_tracker}, Oculus Rift \cite{Rift} and PlayStation VR \cite{PlayStationVR} track the head, hand controllers and other limb-borne accessories using external sensor base stations. Any un-sensed joints can be roughly estimated with inverse kinematics \cite{root_motion_solver, parger2018human}. Other non-worn, external approaches for pose estimation include capacitive sensing \cite{zhang2018wall++}, magnetic fields \cite{Polhemus, trakSTAR}, RF \cite{zhao2018through}, and mechanical linkages \cite{sutherland1968head}.

\subsection{Body Capture with non-IMU Worn Sensors}

Pose estimation using body-worn sensors is more portable and flexible than systems requiring external components. Much research has focused on capturing specific body parts. For instance, hand pose is of great importance in VR/AR applications, and has been tracked using e.g., wrist-worn cameras \cite{kim2012digits,DBLP:conf/sui/ArakawaMKI20,wu2020back}, electrical impedance tomography \cite{zhang2015tomo}, electromyography~\cite{emg1}, depth sensors~\cite{discoband}, magnetic trackers \cite{chen2016finexus}, and specialized gloves \cite{glauser2019interactive}. Faces are important too, most often captured using cameras  \cite{ahuja2018eyespyvr,thies2018facevr}, but other methods such as ultrasound \cite{iravantchi2019interferi} and electromyography \cite{Emg2Gruebler2014DesignOA} have also been explored. 

Our research is more concerned with \textit{whole-body} pose estimation. Many body-worn sensing approaches exist, ranging from exoskeletons \cite{METAmotion}, ultrasonic beacons \cite{vlasic2007practical}, pressure sensors \cite{yin2003footsee} and RFID \cite{jin2018rf}. Body-worn camera approaches are particularly popular, such as work by Shiratori~\etal\cite{shiratori2011motion}, Ng~\etal\cite{ng2020you2me}, and Ahuja~\etal\cite{ahuja2020bodyslam,mecapAhuja:2019}. All of the latter approaches require specialized additional hardware that most users do not own. The goal of our work is to bring the flexibility of body-worn pose estimation to users without requiring them to purchase any new devices. One prior work with a similar mantra is Pose-On-The-Go~\cite{ahuja2021pose}, which estimates a user's full-body pose using only the sensors in a smartphone and when held in the hand. However, much of the full-body pose is guessed (rather than tracked) by measuring absolute movement and using an animated, rigged (IK) avatar. 


\subsection{Body Pose Estimation using Worn IMUs}
In this section, we focus on pose systems relying exclusively on worn IMUs, which most closely matches our technical approach (Table~\ref{tab:imu-systems-overview} provides an overview of prior systems). While single IMUs have been used to track individual limbs (such as arm pose in ArmTrack~\cite{shen2016smartwatch}), it is more common to see "fleets" of IMUs distributed across the body (e.g., the popular XSens~\cite{xsens} suit) for full-body pose estimation. Importantly, these setups are homogeneous in terms of IMUs (and thus performance and noise) and tend to use high-quality sensors running at high framerates not typically seen in consumer mobile devices. As we found, the IMUs utilized in Apple's own ecosystem vary by device, and as such, the data varies in quality, noise and framerate. 

Among prior work using many body-worn IMUs, we see Tautges~\etal\cite{tautges2011motion} able to generate visually plausible motion streams using four XSens IMUs. Sparse Inertial Poser~\cite{von2017sparse} and Deep Inertial Poser~\cite{huang2018deep} use optimization and deep learning-based methods for full-body pose estimation using between 6 and 17 body-worn IMUs. Both systems use SMPL~\cite{Loper2015SMPL, Pavlakos2019SMPL-X}, a statistical body model, as their pose output. Approaches such as TransPose \cite{yi2021transpose} or Physical Inertial Poser~\cite{yi2022physical} build on such efforts and provide more accurate representations and better models.
All these works leverage the fact that the employed IMUs have known and calibrated positions, and the same noise profiles. 


\section{Possible Device Combinations}\label{Possible device combos}

Smartphones, smartwatches, and earbuds have different possible body locations. For instance, a smartphone can be stored in the left or right pocket, held in the left or right hand, held to the head (to take a call), or not carried by the user at all (6 possible states). For smartwatches, they are either worn on the left or right wrist or not worn by the user (3 possible states). For earbud-like devices, they can be worn on the head, placed into a charging case and stored in the left or right pocket, or not carried by the user (4 possible states). Although at present, putting earbuds into a charging case generally puts them to sleep, we assume that in the future a firmware update could allow for continuous IMU data streaming, especially given the larger battery in the case.

Fully enumerated, this yields 72 possible device-location combinations. However, we eliminate three combinations where the user is both wearing earbuds and the phone is held to the head (to take a call), as this is not a typical use case. An additional invalid combination is no phone, smartwatch or earbuds, and thus our system would not run at all. This leaves 68 possible arrangements combinations –- 14 combinations have 1 active device, 36 combinations have 2 active devices, and 18 combinations have 3 active devices. 

\begin{figure}[h]
    \centering
    \includegraphics[width=0.95\linewidth]{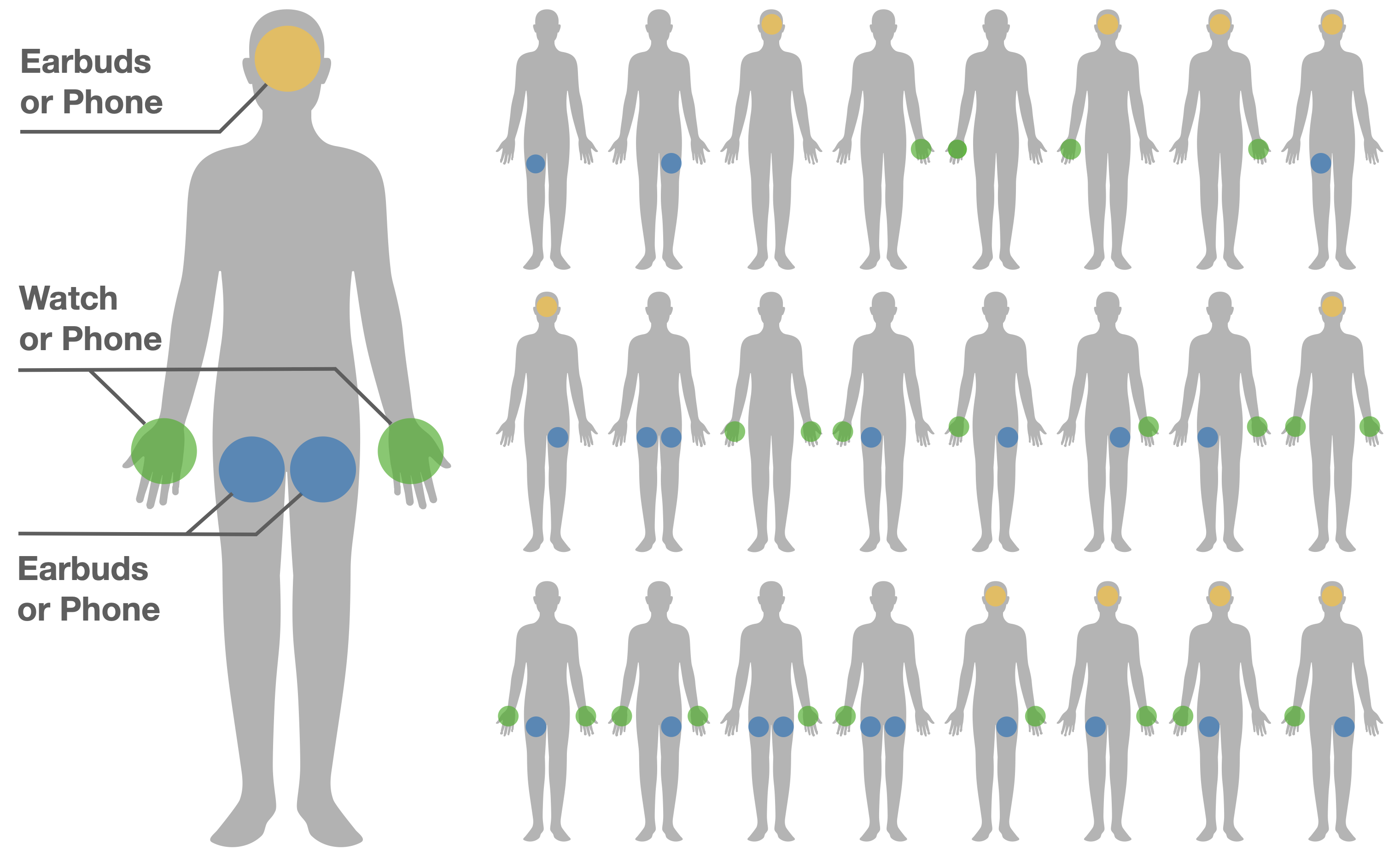}
    \caption{The 24 possible device-location combinations we support and investigate.}
    \label{fig:possible_combinations}
\end{figure}

Next, it is important to consider that some combinations of devices do not provide substantially different body data for our purposes. For instance, a user could wear a smartwatch on their left wrist and hold a smartphone in their left hand -- the IMU data would be highly correlated, and thus we treat it as a single body point. Another example is storing a smartphone in the right pocket, along with earbuds in a charging case -- again, the IMU data would be similar. Thus, what our system truly cares about are the combinations of body location enabled by the 68 possible device-position combinations -- these 24 combinations are illustrated in Figure \ref{fig:possible_combinations}.

We note that our system makes some simplifying assumptions about body positions. For example, in order for a hand position to be considered active, it requires either a smartphone to be held in that hand or a smartwatch worn on that wrist. Even though the signal is not identical, it is highly correlated such that the information power is similar. Similarly, a smartphone held to either ear is considered to be a head location (rather than left or right ear). We made the latter simplification because Apple's AirPods (which we use in our real-time implementation) fuse their IMUs to provide a single-head 6DOF estimate, rather than provide IMU data from each Airpod individually.

\section{Implementation}

\begin{figure}[b]
    \centering
    \includegraphics[width=\linewidth]{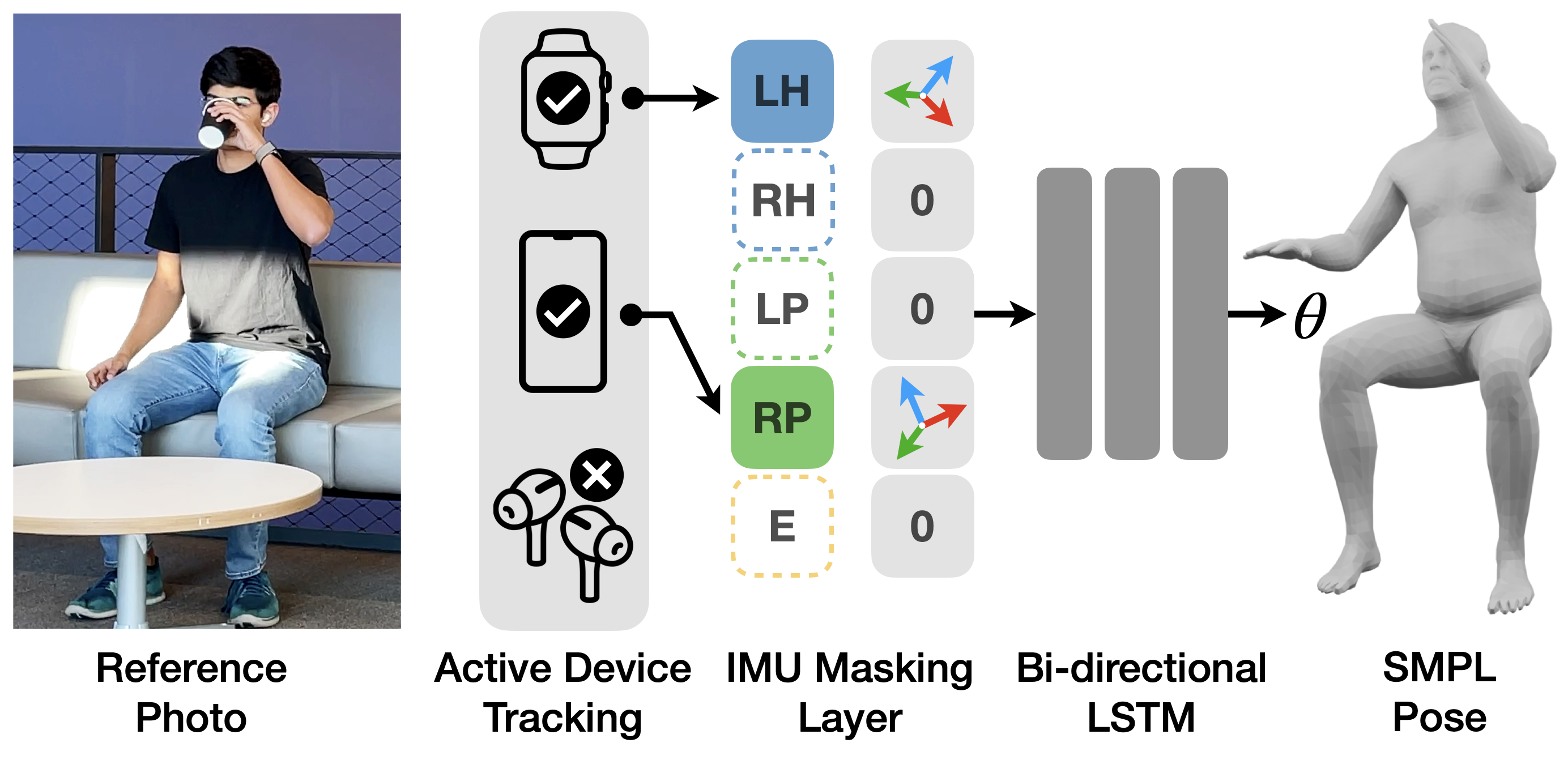}
    \caption{Overview of our real-time system architecture.}
    \label{fig:architecture}
\end{figure}

Figure~\ref{fig:architecture} provides an overview of our pipeline. We focus on three popular consumer devices: smartphone, smartwatch and wireless earbuds/headphones. Each of these devices contains an IMU, the ability to wirelessly transmit data, and some local compute. We envision our model executing on the most capable device carried by the user, with the other less-capable devices streaming their IMU data over e.g., Bluetooth. 

\subsection{Model}
\label{ref:model_arch}

For the learning architecture, we use a two-layer Bidirectional LSTM, inspired by prior works \cite{yi2021transpose, huang2018deep}. Although we did experiment with newer architectures such as transformers, we found these models did not perform well in practice. LSTMs produced smoother output predictions than the other models we tested. For each available IMU, our system uses orientation (represented as a 3$\times$3 rotation matrix) as well as acceleration as input, both in a global coordinate frame of reference. In contrast to prior work, we do not normalize these inputs to be relative to a root IMU sensor location, such as the pelvis, as our available devices vary. We flatten and concatenate these inputs to form an input vector of size 60: 5 possible IMU locations $\times$ (3 acceleration axes + 3$\times$3 orientation rotation matrix), which we input to the model. Of note, our model can ingest any subset of the available IMU data, with absent devices masked (i.e., values set to zero). 


The input vector is first transformed into an embedding of dimension 512 using a ReLU \cite{nair2010rectified} activated linear layer. Next, these embeddings are fed sequentially into a Bidirectional LSTM of hidden dimension 512. A final linear layer outputs 144 SMPL \cite{Loper2015SMPL} pose parameters --- 24 joints represented as 6D rotations (which is a smoother representation space \cite{6drotHaoLi2018}). SMPL also provides a body mesh (6890 vertices), which can be seen in Figures \ref{fig:teaser}, \ref{fig:architecture} and \ref{fig:sample_mesh}. In total, our neural network model has $10.7$M trainable parameters. During training, a forward kinematics module calculates joint positions from these pose parameters and further minimizes it with respect to the ground truth.




\subsection{IMU Dataset Synthesis}\label{ref:dataSynthsec}

To train our pose model, we required a significant volume of data. For this, we can leverage existing motion capture databases to generate a large synthetic corpus. Specifically, we use AMASS~\cite{mahmood2019amass}, a compilation of 24 motion capture datasets (ACCAD, BioMotion, CMU MoCap, MIXAMO, Human Eva, Human 3.6M, etc.) totaling almost 63 hours of high-quality, high-resolution motion capture data in the SMPL~\cite{Loper2015SMPL} format. A wide variety of motions and activity contexts are included, such as locomotion, sports, dancing, exercising, cooking, and freestyle interactions. For additional details on the composition of the AMASS dataset, please refer to~\cite{mahmood2019amass}. 
We note that AMASS has been used in much prior work \cite{huang2018deep, yi2021transpose, yi2022physical} as the basis for deriving synthetic datasets.

The consumer devices we use for our study and real-time demo (described in Sections \ref{sec:data_collection_apparatus} and \ref{ref:realtimeimp}) run at a common framerate of 25 FPS. Thus we resample AMASS' 60$\sim$120 FPS data to 25 FPS. We then follow the synthetic data generation process used in TransPose \cite{yi2021transpose} and DIP \cite{huang2018deep}. In short, we "attach" virtual IMUs to specific vertices in the SMPL mesh (the left and right wrists, the left and right front pant pockets, and the scalp) and compute synthetic acceleration data using adjacent frames in the global frame of reference. To generate synthetic orientation data, we calculate joint rotations relative to the global frame by compounding local rotations starting from the joint to the pelvis (root) following the SMPL kinematic chain. We scale acceleration data ($m/s^2$) by 30 to be suitable for neural networks~\cite{yi2021transpose}. Finally, rather than adding synthetic high-frequency noise to our dataset, we instead smooth both synthetic and real-world data using an averaging window of length $5$ frames ($200$ ms), similar to \cite{transformerIntertialPoser2022}.




We use this pipeline to create 24 sets of data, one for each of our 24 device-location combinations (Figure \ref{fig:possible_combinations}), which we combine into a single dataset. We simulate missing devices by masking-out (i.e., zeroing-out) IMU data for those locations. For example, even in our best-case scenario of three devices present, this means that 2/5$^{ths}$ of the input vector is null. 63 hours of AMASS data $\times$ 25 FPS $\times$ 24 device-location combinations yields 134.8M synthetic IMU instances  with paired ground truth SMPL poses for training. 

\subsection{Training}

The model is trained end-to-end using PyTorch and PyTorch Lightning deep learning frameworks. We use a batch size of $256$ and update the weights using the Adam optimizer with a learning rate of $3e^{-4}$. While training, we use non-overlapping windows of paired IMU and pose data in 5-second (125 samples) chunks. As mentioned earlier, we train our model to regress to full-body pose and full-body joint positions using mean squared error (MSE) loss. Our total loss is the sum of these two individual losses. We train our model for 80 epochs ($22$ hours) on an NVIDIA Titan X GPU.

\subsection{Joint Rotation Refinement}

As the last step of our inference pipeline, we adopt the Inverse Kinematic refinement method presented in \cite{jiang2022avatarposer} to perform a final refinement of our output pose. Although our model predicts the rotation of legs, hands and head, it does not necessarily fully honor the absolute orientation offered by the IMUs, even when weighted heavily in our loss term. However, it is logical to take advantage of IMU orientation for limbs with devices, as it is both an absolute value and considerably less noisy than accelerometer data. More specifically, as we have absolute orientation from the IMUs, we optimize certain bone orientations for each instrumented joint. In particular, for the wrist joint (smartwatch/phone), we optimize the elbow and the shoulder orientations, and similarly for the head (earbuds/phone) and hip (smartphone/earbuds case) joints. We implement this using the PyTorch framework and optimize this error using the MSE loss and the Adam optimizer. We allow this optimization to run for 10 iterations on each frame, which we found to not impede real-time performance.
\begin{figure}[b]
    \centering
    \includegraphics[width=\linewidth]{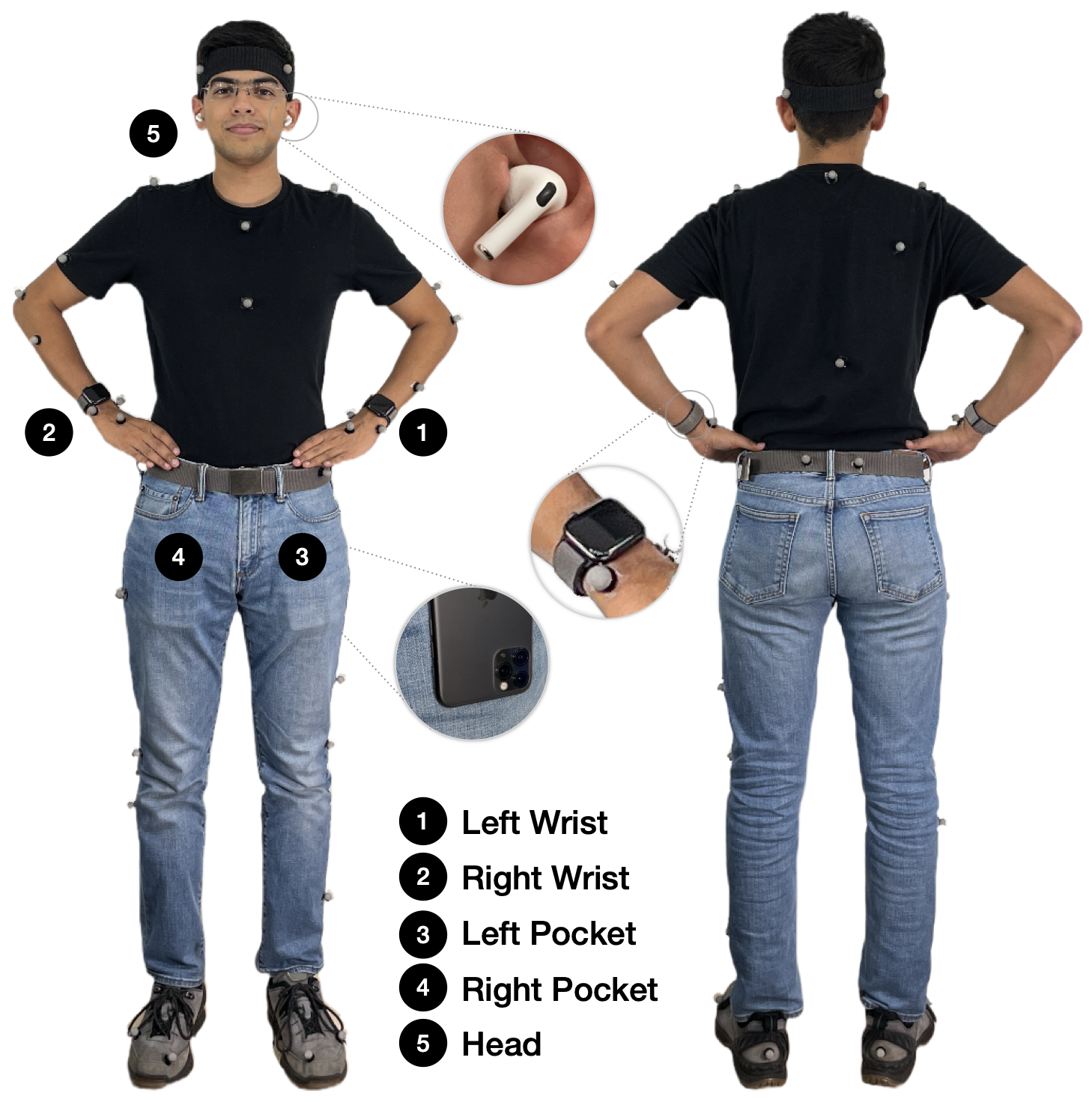}
    \caption{IMUPoser data collection setup. Participants wore 2 smartwatches, kept 2 smartphones in their front pockets, and wore wireless earbuds. 41 retroreflective motion capture markers were also placed around the body to track ground truth body pose.}
    \label{fig:ipose_data_collection_setup}
\end{figure}

\section{Evaluation}
We systematically isolate and analyze the efficacy of IMUPoser across different datasets and conditions. 

\subsection{DIP-IMU Dataset}
To test the performance of our model on real (and not synthetic) IMU data, we use DIP-IMU \cite{huang2018deep}, an IMU-based MoCap dataset. While smaller than the AMASS dataset we used for training, it offers a good variety of poses and activities across five classes: upper-body (arm raises, stretches, swings, etc.), lower-body (leg raises, squats, lunges, etc.), interaction (gestures to interact with everyday objects), freestyle (jumping jacks, punching, kicking, etc.) and locomotion (walking, side steps, etc.). A secondary benefit of using DIP-IMU is that it has been used for evaluation in other similar works \cite{yi2021transpose,huang2018deep,yi2022physical,transformerIntertialPoser2022}, permitting direct comparison. DIP-IMU used the commercially-available Xsens \cite{xsens} IMU-based system to capture data from 10 participants. The data is sampled at 60~Hz, leading to a total dataset size of approximately 90 mins.

\subsection{IMUPoser Dataset}
\begin{figure*}[t]
    \centering
    \includegraphics[width=\linewidth]{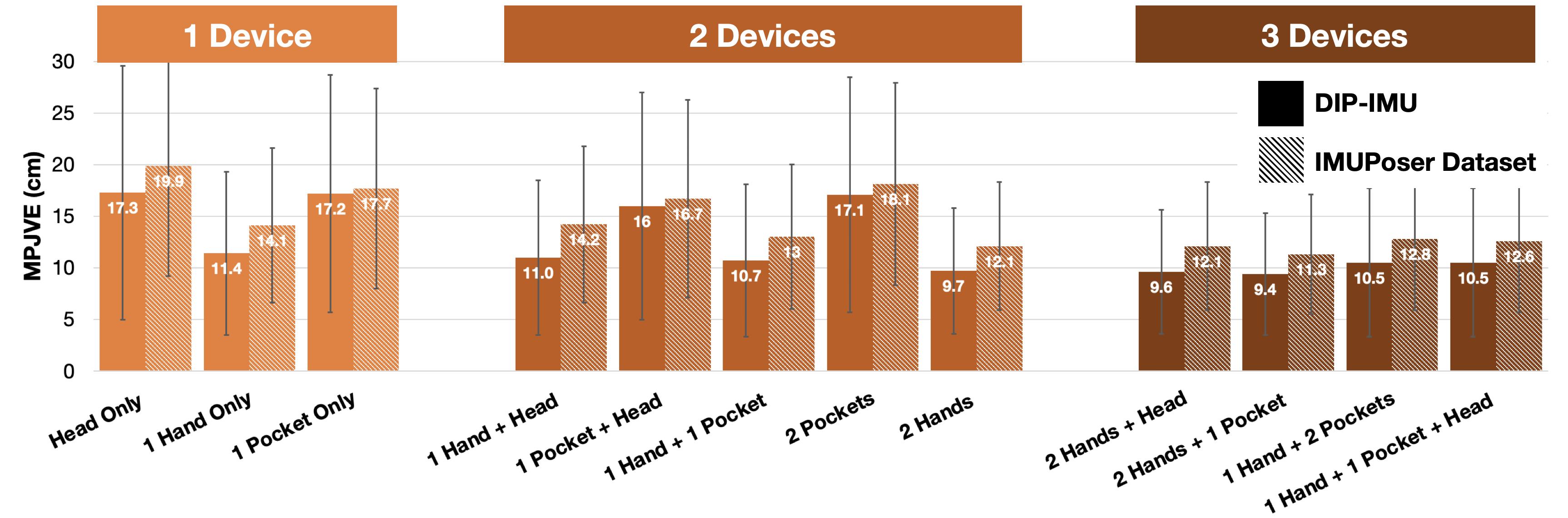}
    \caption{Accuracy across different device combinations. Error is Mean Per Joint Vertex Error (MPJVE) in cm. Note how error decreases as the number of devices increases.}
    \label{fig:sample_device_acc}
\end{figure*}

As noted above, DIP-IMU used the professional-grade XSens system for data collection, which costs approximately \$4000 USD. All of the IMUs are matched, offering similar noise and tracking performance. To complement this dataset with a \textit{consumer} device equivalent, we collected our own dataset. 

\subsubsection{Data Collection Apparatus}
\label{sec:data_collection_apparatus}
Our data collection apparatus consisted of two smartphones (Apple iPhone 11 Pro) placed in the left and right front pockets, two smartwatches (Apple Watch Series 6) placed on the left and right wrists, and one pair of Apple AirPods Pro worn in the ears (Figure \ref{fig:ipose_data_collection_setup}). The sampling rate of our system was configured to 25~Hz, the maximum sampling rate of the AirPods. The Apple Watch and AirPods communicated over Bluetooth to the iPhones, and the two iPhones relayed all IMU data to a laptop for data processing and recording. Although users had all five devices on them during data capture, we only use a subset of these devices for pose estimation, as described in Section \ref{Possible device combos}. 

For ground truth pose, we use a Vicon Motion Capture System system~\cite{Vicon} with twelve MX40 cameras and four T160 cameras capturing at 120FPS. We used Vicon Blade 3.2 for capture and data export and Vicon IQ 2.5 for data cleaning. We downsample the Vicon data and synchronize it with our collected IMU data streams. For analysis, we fit an SMPL mesh to the Vicon data using Mosh++~\cite{mahmood2019amass}. 


\subsubsection{Device Calibration} \label{Calibration}
In contrast to commercial IMU-based motion capture systems like XSens, smartphones, smartwatches, and earbuds fail to provide IMU orientations in a common (global) frame of reference. If a device contains a magnetometer, the manufacturer usually provides a way to access the orientation of the device in a global frame of reference oriented with Earth's gravitation and magnetic fields. While the iPhones and Apple Watches that we used for this study contained magnetometers, we found their global orientation data to be fairly noisy. Moreover, the Apple AirPods do not contain magnetometers and hence only provide orientation relative to the initial frame of reference of the head. As a result, we opted to use the XArbirtraryCorrectedZVertical frame of reference provided by the Swift CoreMotion API \cite{CMAttitudeRef}. 

Before the study began, we aligned all the devices to a common frame of reference and recorded their orientation values over a window of three seconds. This acted as calibration data, bringing all the devices into the same global frame of reference. In practice, since the AirPods only sampled IMU data when they were in a participant's ears, the common frame of reference was set to that. In line with prior works \cite{yi2021transpose, huang2018deep}, we asked participants to make a T-pose for three seconds to calculate the orientation offsets between the device and the bone joint that it was attached to. The T-pose acts as a template pose wherein rotations are identity and thus known for each joint. This helps calibrate for users wearing the devices in different orientations, for example, a phone held in the hand vs. a watch worn on the wrist.

\subsubsection{Data Collection Procedure}
\begin{figure*}[b]
    \centering
    \includegraphics[width=\linewidth]{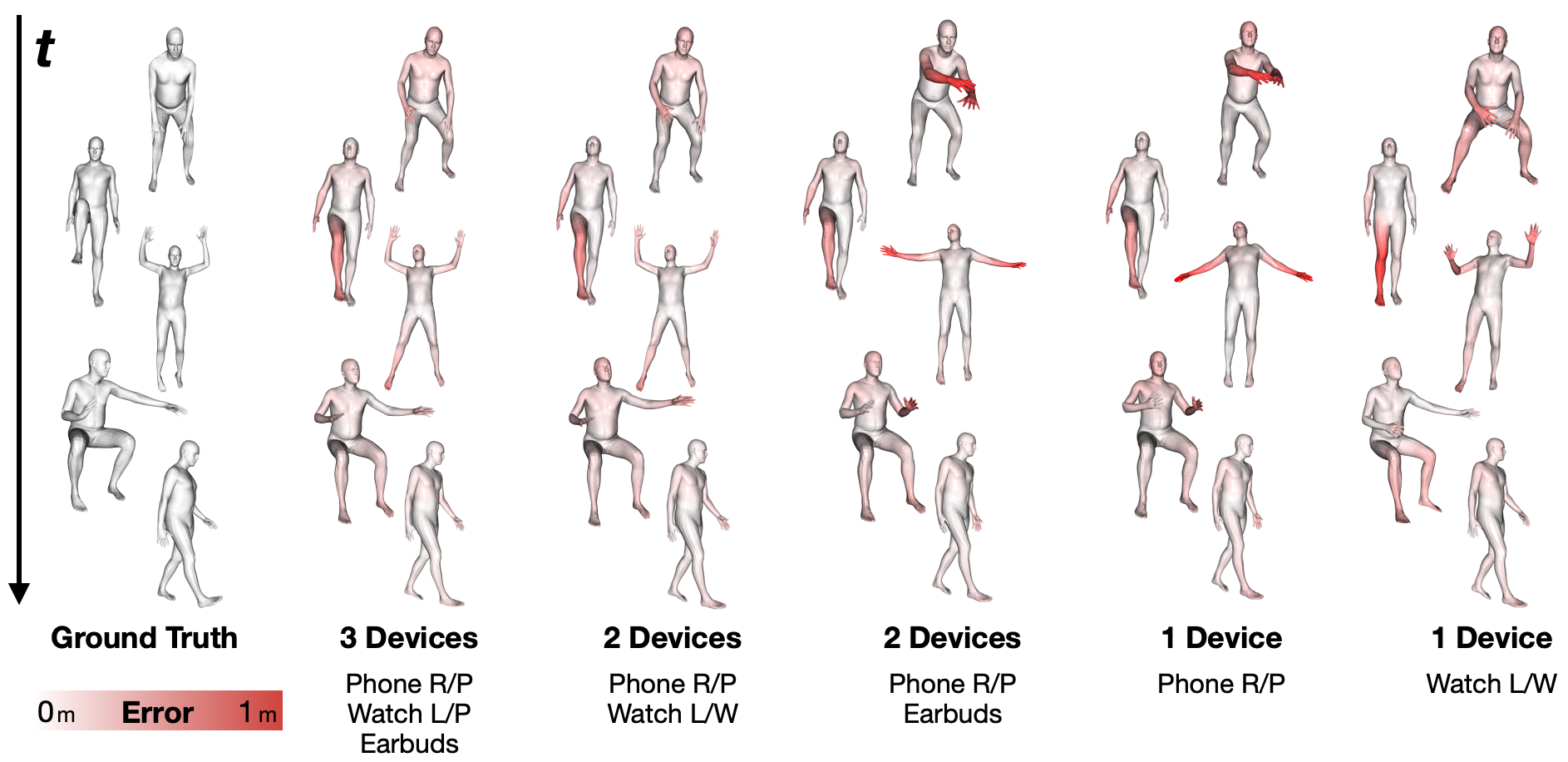}
    \caption{Sample SMPL mesh predictions for different device placements and combinations. The red color indicates the per vertex error in meters (ranging from 0 to 1 m). }
    \label{fig:sample_mesh}
\end{figure*}

For our data collection, we recruited 10 participants (5 identified as female, 5 identified as male) with a mean age of 22. The study lasted roughly 45 minutes and paid \$20 in compensation. We asked participants to wear and store the five devices in the way that felt most natural to them. Other than requesting the participants to wear pants with pockets, we did not control for differences in clothing, pocket styles, or smartwatch placement preference on the wrist, so as to get realistic real-world variation. For our Vicon-derived ground truth, we placed 41 optical markers on participants. In order to keep the markers secure, we asked participants to tuck in their shirts and provided velcro straps where needed.

Inspired by prior works \cite{huang2018deep,ahuja2021pose}, we collected our data using an "obstacle course"-style procedure. We extended the classes in the DIP-IMU dataset and included the following motions:

\begin{itemize}[leftmargin=*]
    \item \textit{Upper Body}: Right arm raises, left arm raises, both arm raises, right arm swings, left arm swings, both arms swinging, arms crossing across the torso, and arms crossing behind the head. 
    \item \textit{Lower Body}: Right leg raises, left leg raises, squats, lunges with left leg, and lunges with right leg.
    \item \textit{Locomotion}: Walking in a straight line, walking in a figure 8, walking in a circle, sidesteps with legs crossed, and sidesteps with feet touching.
    \item \textit{Freestyle}: Jumping jacks, tennis swings, boxing with alternate arms, kicking with the dominant leg, push-ups, and dribbling a basketball.
    \item \textit{Head Motions}: Moving head up-and-down, moving head left-to-right, leaning head from shoulder-to-shoulder, and moving head in circles.
    \item \textit{Interaction}: Scrolling on a smartphone while seated in a chair. 
    \item \textit{Miscellaneous}: Waving with right arm, waving with left arm, clapping, hopping on right leg, hopping on left leg, jogging in a straight line, and jogging in a circle.
\end{itemize}

The upper body, lower body, locomotion, freestyle, head motions, interaction and miscellaneous scenarios lasted for 69.7, 43.4, 95.3, 76.2, 36.8, 19.2, and 74.3 seconds on average, respectively, resulting in roughly 7 minutes of data per participant. All the motions were continuous and data was also captured while participants were transitioning from one category to another. 

\subsection{Evaluation Protocol}
In order to compare with prior works, we follow the exact method detailed in  \cite{huang2018deep,yi2021transpose,transformerIntertialPoser2022,yi2022physical}. Specifically, we use data from the first eight participants of DIP-IMU as training data, with the last two participants used for testing. We fine-tune our AMASS-trained model using this training data downsampled to 25~Hz to match both our AMASS training data and our real-time system's capabilities. We further test this model on our IMUPoser Dataset, helping to assess real-world accuracy and performance. Our model is evaluated in an online fashion. In particular, we feed a rolling window of 125 samples (5-second history) with a 1-sample overlap, emulating real-world use. This data is smoothed using an averaging filter, as described in Section \ref{ref:dataSynthsec}. We analyze these results using different evaluation metrics across various device-location combinations. Also following prior work \cite{yi2021transpose, yi2022physical, huang2018deep}, we make use of the following evaluation metrics to quantify the performance of our full-body pose estimation pipeline:

\begin{enumerate}[leftmargin=*]

  \item \textit{Mean Per Joint Rotation Error:} MPJRE measures the mean global angular error across all joints in degrees (\textdegree). 
  
  \item \textit{Mean Per Joint Position Error:} MPJPE measures the mean Euclidean distance error of all estimated joints in centimeters (cm) with the root joint (pelvis) aligned. 
  
  \item \textit{Mean Per Joint Vertex Error:}  MPJVE measures the mean Euclidean distance error across all vertices of the estimated SMPL mesh in centimeters (cm) with the root joint (pelvis) aligned. 
 
  \item \textit{Mean Per Joint Jitter (Jitter):}  Jitter measures the average jerk of the predicted motion \cite{yi2021transpose}. A lower jerk value signifies a smoother and more natural motion.
  
\end{enumerate}

We use mesh error (MPJVE) as our primary evaluation metric for most tasks, due to its ease of understanding and its utility as a benchmark for comparison with prior work.

\section{Results}
\begin{figure*}[b]
    \centering
    \includegraphics[width=\linewidth]{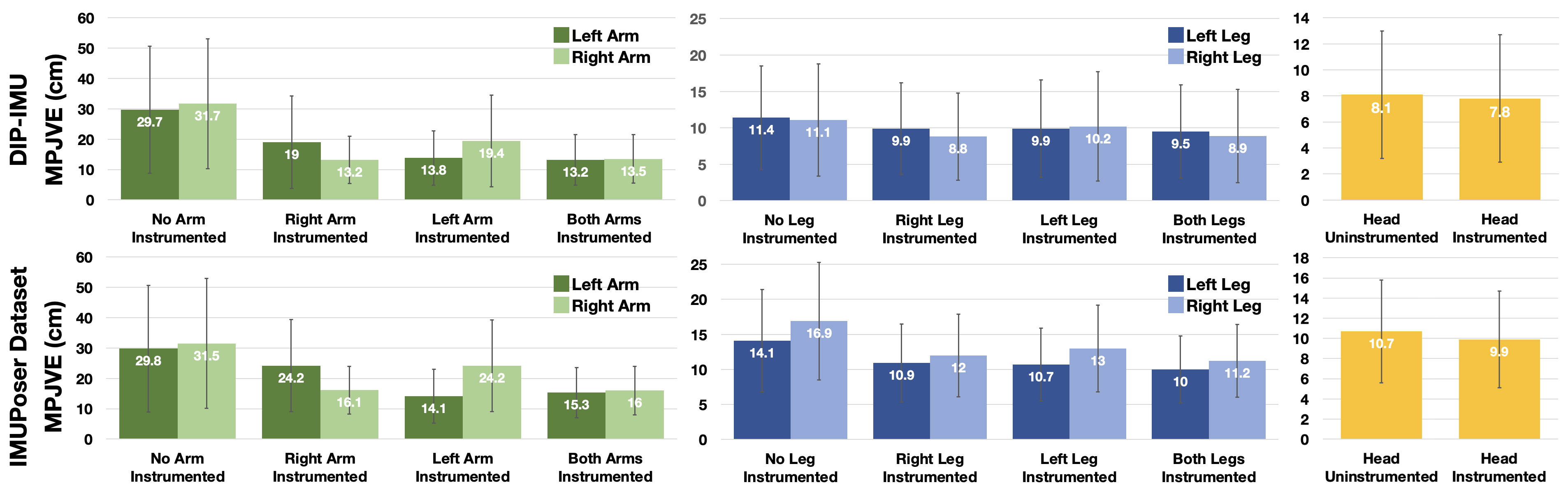}
    \caption{Summarized accuracy results across different body regions evaluated on the DIP-IMU and IMUPoser datasets.}
    \label{fig:body_region}
\end{figure*}

We first describe IMUPoser's accuracy across device-location combinations, before changing our focus to look at results by body region. We conclude this section with a comparison to other related systems. 

\subsection{Accuracy Across Device-Location Combinations}

To simplify presentation of results, we group the 24 possible device-location combinations (Figure~\ref{fig:possible_combinations}) into 12 supersets based on the number of devices present and their body locations (ignoring left/right placements). Figure~\ref{fig:sample_device_acc} presents the results for the IMUPoser and DIP-IMU datasets. Note, that our model has not been fine-tuned on the IMUPoser Dataset. Across all device combinations, we find a MPJVE of 14.1~cm on the IMUPoser Dataset and a MPJVE of 12.1~cm with DIP-IMU. When averaging the results across both datasets, having one device on the user results in a MPJVE of 16.27~cm (SD=9.93~cm), which decreases to 13.9~cm (SD=8.36~cm) when a second device is present. The lowest error, unsurprisingly, is when three devices are present -- a MPJVE of 11.1~cm (SD=6.51~cm) across all possible three-device combinations.


Figure~\ref{fig:sample_mesh} offers example mesh predictions across different device-location combinations. As expected, accurate head orientation estimation is only plausible when earbuds are present. Other times the head regresses to the most natural orientation given where the body is facing. Global body orientation works best when at least two devices are present. Lastly, motions that have a characteristic cadence, such as walking, work well across all combinations. Similarly, activities with symmetric limb motions, such as jumping jacks, work fairly well even with no sensor data from important limbs. On the other hand, activities with uncorrelated limb motions fail unless limbs are instrumented. 

\subsection{Accuracy Across Body Regions}

Figure~\ref{fig:body_region} provides a breakdown of system accuracy across different body regions for the IMUPoser and DIP-IMU datasets. We note that the accuracy for a limb with an instrumented point is always greater than that of an uninstrumented one. For example, averaging across both datasets and with an IMU present on the right hand, the MPJVE is 14.65~cm for the right arm (right hand = 17.2~cm)  vs. 21.6~cm for the left arm (left hand = 26.9~cm). Unsurprisingly, the highest error is when none of the limbs in a particular body region have IMU data.

Also unsurprising is that the lowest error is achieved when both left and right limbs have IMUs present. For example, with only one IMU on the arms, the MPJVE for both arms is 18~cm (right hand = 22.17~cm; left hand = 20.4~cm). Whereas with both arms having IMUs, the MPJVE is 14.5~cm (right hand = 17.35~cm; left hand = 16.9~cm). A partial exception to this trend is the legs. Unlike the arms, which can move independently, legs tend to move in tandem (out of phase when walking, or in phase for activities such as jumping). This means that even one IMU on the legs is still highly effective at predicting both legs, and two IMUs offer just a modest gain. Looking at our results, the MPJVE for the left leg is 10.3~cm (left foot = 15.65~cm) when the IMU is in the left pocket, and the error for the right leg is 10.4~cm (right foot = 16~cm) when the IMU is in the right pocket. When both IMUs are present (i.e., left and right pockets), the error of the left and right legs drop very modestly to 10.05~cm and 9.75~cm, respectively. 

We note that error accumulates along the kinematic chain (see Figure \ref{fig:sample_mesh}). Across all conditions, the average error of the end-effectors (left hand, right hand, left foot, right foot, head) is 20.28~cm and 17.29~cm on the IMUPoser Dataset and DIP-IMU Dataset, respectively (vs. 12.92~cm and 11.23~cm for joints that are not end-effectors). 

\subsection{Comparison to Prior Work}
\begin{table*}[t]
\centering
\begin{tabular}{l c c c c c}
\toprule
\textbf{System} & \textbf{\# Inst. Joints}    & \textbf{MPJRE (\textdegree)}  & \textbf{MPJPE (cm)} & \textbf{MPJVE (cm)} & \textbf{Jitter ($10^2m/s^3$)}\\
\midrule

SIP (offline)   & 6 & 8.7  & 6.7 & 7.7 & 3.8 \\
DIP (online)   & 6 & 15.1  & 7.3 & 8.9 & 30.13 \\
TransPose (online)  & 6 &  8.8 & 5.9 & 7.1 & 1.4 \\
PIP (online)  & 6   & -  & - & 5.9 & 0.24 \\

\rowcolor{Gray}
IMUPoser (online)  & 1--3  & 23.9  & 9.7 & 12.1 & 1.9 \\

\bottomrule
\end{tabular}
\vspace*{2mm}\caption{Comparison of IMUPoser to key prior work, all evaluated on the DIP-IMU Dataset \cite{huang2018deep}.}
\label{tab:comparison}
\end{table*}
To the best of our knowledge, no prior research has investigated deriving full-body pose from such a sparse set of consumer-grade devices equipped with IMUs. Table~\ref{tab:comparison} offers a quantitative comparison against key prior work, all evaluated on the same DIP-IMU Dataset \cite{huang2018deep}. 

Unsurprisingly, for a system that uses between 1 and 3 IMUs, our model is less accurate than those utilizing 6 sensors (i.e., IMUs placed on each limb). However, compared to DIP \cite{huang2018deep} and Transpose \cite{yi2021transpose}, our MPJVE is only worse by 3.2~cm and 5.0~cm, respectively. It is interesting to note that the Jitter of our system is in line with prior work (1.9 vs. 1.4 of TransPose). At a high level, even with an impoverished sensing configuration, we are able to produce natural, realistic and smooth pose estimation sequences. In the future, we hope to combine physics-backed models (as in PIP) to further improve the pose estimation of our system. 

\section{Active Device Tracking}
A crucial piece of information our pose model needs before it can run is: 1) What devices are present on the user? And 2) where these devices are located on the body? For this, we created a separate piece of software, which runs in parallel with our pose model. 


\subsection{Implementation}
To determine where devices are located on the body, we require three pieces of information from the user, which we envision being collected when a user first purchases a device. 1) In which pocket do they typically store their phone? 2) In which hand do they typically hold their phone? And 3) On which arm would they wear a smartwatch? After this basic initialization, we use a series of automated heuristics.


We make the assumption a smartphone is held in the hand if the screen is on and the IMU is reporting even slight motion. If the user is wearing a smartwatch, we can use the distance between the watch and phone (provided by Apple's NINearbyObject API \cite{NINearbyObject}, which uses UWB) to guess the holding hand automatically (see Figure~\ref{fig:active_device}). If no smartwatch is worn, our system falls back on the hand specified by the user during setup. If the smartphone screen is off and the IR proximity sensor is triggered, we assume the phone is in a pocket. If the user has a smartwatch, we can similarly use UWB-derived distance to guess the pocket. If no smartwatch is worn, we default to the user-specified pocket.

As most users wear watches in a consistent location, the logic for smartwatches is simpler. If it is connected to the iPhone and moving, we assume it is worn on the user-specified hand. Similarly, for Airpods, if they are connected to the iPhone, we know they are in the ear. 
When in their charging case, Airpods go to sleep and stop transmitting IMU data. However, we believe that Apple could modify the Airpods firmware such that in the future they could continue to transmit IMU data even when stored in a pocket.
\subsection{Evaluation}

\begin{figure}[h]
    \centering
    \includegraphics[width=\linewidth]{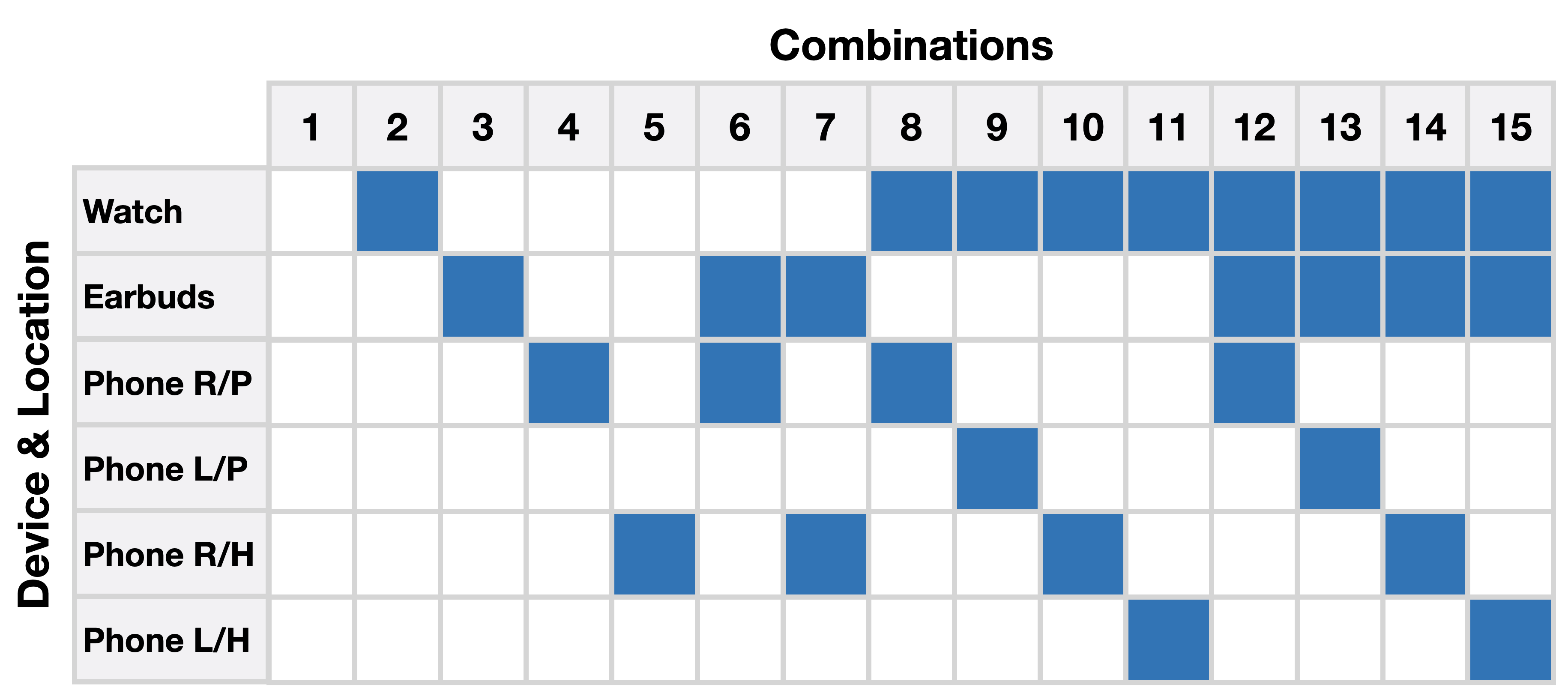}
    \caption{Device combinations tested as part of our active device tracking study. Blue denotes presence in the set. }
    \label{fig:active_device_combos}
\end{figure}

\begin{figure*}[t]
    \centering
    \includegraphics[width=\linewidth]{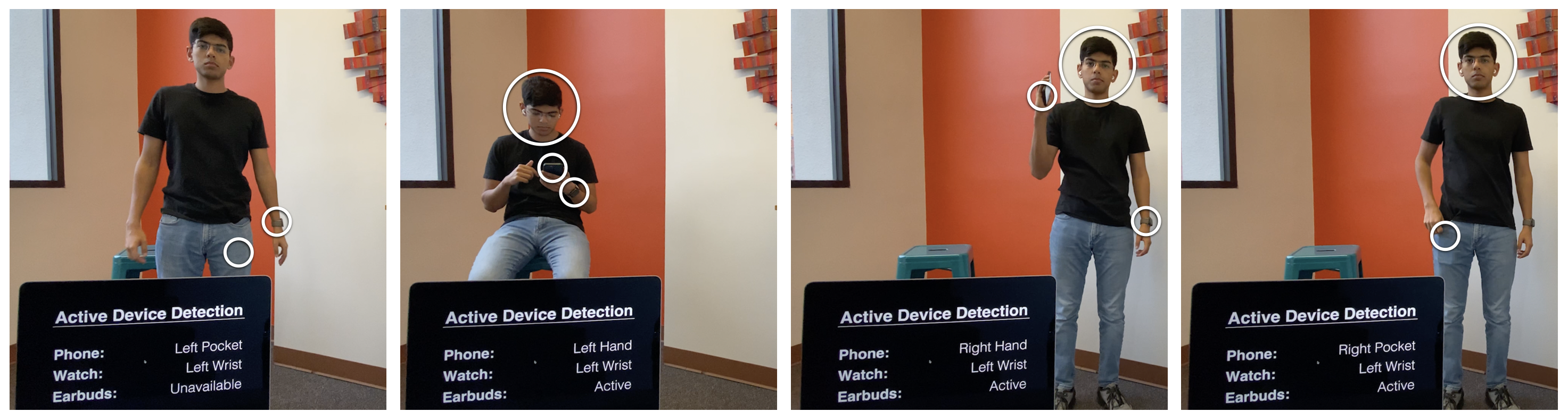}
    \caption{Active device tracking across different device combinations. Active devices are highlighted with a white circle for illustration and the foreground laptop shows the live tracking result.}
    \label{fig:active_device}
\end{figure*}

As a preliminary evaluation of our active device tracking prediction, we ran a user study with 7 participants (5 identified as male, 2 identified as female; mean age 27.8; all right-handed with a preference for wearing watches on the left wrist). The study lasted approximately 15 minutes and paid \$5. To initialize our system, we recorded participants' answers to the three preference questions listed in the previous section. We then asked users to transition between 15 device combinations, in a random order, documented in Figure \ref{fig:active_device_combos}. When a device was not in a requested set, it was set aside on a nearby table. For each requested device-location combination, we asked participants to walk around for about 10 seconds, then sit down briefly, rise to stand again, and lastly return to the starting position. Before the next trial began, the necessary devices were given or taken from the participants. Throughout the study, our active device tracking process ran, making live predictions about what devices were active and where they were located. A trained experimenter conducted the study, marking the start and stop of each device combination trial, alongside ground truth labels. 

Across all participants and all data instances, the accuracy of earbuds and smartwatch tracking was 100\%, owing to their known locations and very reliable detection of worn vs. not worn. Smartphone tracking is the most challenging, with five possible states (not present, left pocket, right pocket, left hand, right hand). We found the instance-wise accuracy for smartphone tracking was 90.8\%. 


\section{Real-Time Implementation}
\label{ref:realtimeimp}

To help demonstrate the imminent feasibility of our approach, we created a real-time implementation of IMUPoser, which can be seen in our Video Figure. It is comprised of two main processes working together. First is active device tracking, which monitors what devices are available to provide IMU data and predicts where they are located on the body. Second is our pose model, which is passed the location inferences and IMU data. 

\subsection{Proof-of-Concept Device Ecosystem}
As a proof-of-concept implementation, we use an Apple iPhone 11 Pro, Apple Watch Series 6, and AirPods Pro. Apple offers a mature inter-device API that allows these devices to exchange data. Each device reports 6DOF IMU data at different rates, with the slowest being AirPods at roughly 25 FPS. We also note that although AirPods come as a pair, they fuse their individual IMUs into a single 6DOF head estimate.

\subsection{Output}\label{real time pose prediction}
As a proof of concept, we use an iPhone optionally connected to an Apple Watch and AirPods. The iPhone streams all available IMU data back to a MacBook Air (2021), which runs our active device tracking and pose estimation processes, with a mean inference time of 26.8~ms. We believe our model could be run on a mobile phone with additional engineering effort. Regardless of where the model runs, it is capped at 25 FPS, the reporting frequency of our slowest IMU (Airpods). Before running our system, we must perform the same calibration as in data collection (see Section \ref{Calibration}). For real-time output, we visualize the SMPL mesh.

\section{Open Source}

To enable other researchers and practitioners to build upon our system, we have made our dataset, architecture, trained models, and visualization tools freely available at \url{https://github.com/FIGLAB/IMUPoser} with the gracious permission of our participants.

\section{Limitations and Future Work}

While IMUPoser enables pathways to full-body pose estimation with minimal user instrumentation, it has pros and cons like any other technical approach. While IMUPoser can glean insights about the pose of limbs for which it has no direct sensor data, it is important to note that such a pose is only an approximate result. For cases where the motion of the instrumented joint is completely independent of that of the uninstrumented one, IMUPoser tends to regress to the mean pose. IMUPoser can support the incorporation of new joint locations by using the corresponding SMPL mesh vertices for training. Thus, in the future, IMUPoser can potentially support and track new device placements, such as a phone in a back pants pocket, coat pocket, armband, etc. The fidelity of the system could also be improved by integrating additional consumer devices (e.g., smart shoes, eye-wear, rings) into the ecosystem. This would help expand the range of poses supported by IMUPoser, allowing it to track dynamic activities such as cycling, kayaking, skiing, etc. 

Unlike Transpose~\cite{yi2021transpose} and PIP~\cite{yi2022physical}, the current implementation of IMUPoser does not predict global root translation. In the future, using better learning methods and multimodal cues when available (e.g., visual odometry from the smartphone \cite{ahuja2021pose}) could help predict translation. Along similar lines, the overall accuracy of the system could be improved by including contextual cues such as the activity being performed \cite{ahuja2022activityposer} or the user's location.    

Another limitation of our system is active device tracking. Currently, this is a basic, proof-of-concept implementation that needs further refinement before it can be deployed for consumer use. Furthermore, all of the devices need to be in a homogeneous ecosystem (e.g., Apple) to work effectively. In the future, the use of a common industry-wide standard to connect and network between different consumer devices can help mitigate this issue.

Finally, we envision IMUPoser executing on the most capable device the user happens to be carrying. In most cases, this will be a smartphone, but not always, especially in the future. For instance, it is possible today for a user to go for a run with a smartwatch and wireless headphones, but no phone. In the near future, it seems possible there will be AirPod-like devices that can operate independently (e.g., hearables). 





\section{Conclusion}

In this paper, we presented IMUPoser -- a system for real-time, full-body pose estimation using IMUs present in consumer devices such as phones, smartwatches and earbuds. Our system must automatically track devices that are available and where they are currently located on the body, and use streaming IMU data to estimate pose. Our evaluations show that IMUPoser can contend with the noisy signals of consumer IMUs and produce natural and temporally-coherent pose estimates with as little as one device. This opens up new and interesting whole-body applications with no additional user instrumentation. 

\begin{acks}
We would like to thank Justin Macey and the Computer Graphics Lab at Carnegie Mellon University for assisting with collecting and processing the IMUPoser Dataset. We are also grateful to Giorgio Becherini and the Perceiving Systems group at the Max Planck Institute for Intelligent Systems for help with processing our motion capture data.
\end{acks}

\bibliographystyle{ACM-Reference-Format}
\bibliography{sample-base}
\end{document}